\documentclass[a4paper,12pt]{article}

\usepackage[latin2]{inputenc}
\pdfoutput=1
\usepackage{t1enc}
\usepackage{anysize}
\usepackage{amsmath}
\usepackage{amsfonts}
\usepackage{amssymb}
\usepackage{graphicx}
\usepackage{subfigure}
\usepackage{multirow}
\usepackage{booktabs}
\usepackage{listings}
\usepackage{float}
\usepackage{listings}
\usepackage{bm}
\usepackage{url}
\usepackage{authblk}
\usepackage[table,xcdraw]{xcolor}
\usepackage{physics}
\usepackage{fancyvrb}
\usepackage{enumitem}

\begin{document}


\title{Mathematical modelling European temperature data: spatial differences in global warming}

\author[1]{Csilla Hajas}
\author[2]{Andr\'as Zempl\'eni}

\affil[1]{Department of Information Systems, E\"otv\"os Lor\'and University, Budapest, Hungary}
\affil[2]{Department of Probability Theory and Statistics, E\"otv\"os Lor\'and University, Budapest, Hungary}





\maketitle

\begin{abstract}
This paper shows an analysis of the gridded European  precipitation data. We combine simple linear regression with data mining tools like clustering,
and evaluate the strength of the results by the modern bootstrap methods. We have used the 0.5 grade-grid of daily temperatures for 65 years, created by the European Climate Assessment.  We have checked the stability of the results by changing the starting point of the linear regression -- this approach might be valuable for climatologists in finding the
"best" starting point for assessing the global warming in Europe. Different bootstrap approaches were compared and it turned out that the dependent weighted bootstrap is the best for checking the significance of the estimators.
\end{abstract}

{\em MSC codes: 62P12, 62H30, 62F40}

\section{Introduction}  
  
\begin{figure}
		\centering
	\includegraphics[height=70mm]{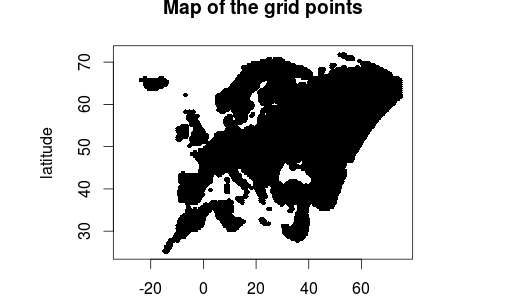} 
	\caption{Map of the region covered by our data} 
	\label{map}
\end{figure}

Temperature changes are in the focus of attention since the global warming became a major threat to the standard of human life. There is a tremendous amount of information available on the subject (see e.g. \cite{gw} and the references therein), showing that the temperature sequences have started to rise from as early as 1960s. There are opinions about the link between NAO (North Atlantic Oscillation) index and low frequency variability of the climate (see e.g. 
\cite{cb}). So as a comparison, 
we have also investigated the residuals, after having removed the effect of the daily NAO index on the temperature time series. This effect was not substantial, but the use of these data allows for checking the robustness of the results. 

The used observations are the 65 years of daily temperature data of the European Climate Assessment from 1950 to 2015 (E-OBS, {\tt {http://www.ecad.eu}}).
We have used the 0.5-grade data, available for Europe and parts of northern Africa.  Figure \ref{map} depicts the covered region. 
This gridded database can be considered as a standard for climate analysis, see \cite{hhgk}. 
Its quality has been evaluated in \cite{hhnj}, and the results show that it may be considered reliable for most of Europe. 
However, especially in the African and Near-Eastern region there are missing periods of various length, which have to be taken into account. The same temperature data set was used in \cite{vz}, where the changes in the bivariate dependence structure were analysed.

Determining the speed of global warming is a very important and actual question, especially if possible confidence intervals are also calculated to the estimators. 
As mathematicians, we cannot give exact explanations for the changes, but we may try to reveal them and to estimate the statistical error of the estimates. 
This estimation is not easy at all, as in the data set there are various types of dependencies, which make the use of standard statistical techniques difficult. 

 The used mathematical models are first the linear regression -- not only for the whole, standardized data set, but for altogether 30 shorter data sets, which were got by omitting the first years sequentially (i.e. the $k$th set consists of the years $(1950+k,\dots , 2015)$. The aim of this approach is to test if there is a tendency in the speed of the warming. 
Our results support the hypothesis of a general increase in the speed of warming, but we have been also interested in the spatial patterns of the processes. 

Thus we applied the Gaussian model-based clustering, see \cite{mbc} for the sequence of estimated coefficients. The proposed method to find the number of clusters is an information criterion (BIC). For assessing the reliability of the results we investigate different bootstrap approaches. We have to
take care on the dependence in the data -- the traditional solutions to this phenomenon are covered in the book of \cite{lah}. However, for the case of
regression models, a new bootstrap 
method has been developed by \cite{wu}. A more recent approach by  \cite{sh} is suitable for the case of dependent residuals -- this is the one we used for assessing the reliability of the results and the significance of the change in speed of the temperature increase.

We give details of the used models in Section \ref{models}. Section \ref{appuni} is devoted to a simulation study on the bootstrap methods.
Section \ref{appl}  is about our results for the modelling of the temperature data. In Section \ref{conclu} we formulate the conclusions.


\section {Methods}
\label{models}

In this section we briefly introduce the used methods. There is no need to introduce the linear models, as they are well-known to everyone. However, in order to assess
the accuracy of the results we got by them, we needed special tools, as the data is both serially and spatially dependent. These will be explained in some detail below.

\subsection{Clustering}
\label{Ch21}

Our data is spatio-temporal. Spatio-temporal data mining has been developed in the last few years, see for example  \cite{she} or  \cite{gia}. 
However, we prefer to use a simple, yet suitable and classical data mining tool, clustering. Here we applied two different methods: density-based ones and the $k$-means clustering. This latter is a traditional, simple and quick method even in our case of over 20000 data points in the 30 dimensional space. We shall see that a practical preprocessing makes this tool especially useful. In our case this approach ensured that we focused on the temporal aspect in the data mining (as the time series of estimated quantiles for the moving windows is analysed), but the method turned out to be suitable for detecting areas sharing similar properties -- thus also the spatial aspects were included in the analysis. The first idea for spatial clustering is to apply the density-based methods. However, it turned out that we have not got any valuable results by these methods. 

When turning to the $k$-means clustering, it is a non trivial question, how to determine the number of clusters. One may use the traditional elbow-rule, based on the portion of the explained variance -- but it usually gives no clear-cut answer. However, the model-based clustering overcomes this problem, as here an adapted version of the Bayesian Information Criterion (BIC) may be applied successfully, see  \cite{mbc}. The method is implemented in the {\tt mclust} package of R. It is assumed  that the observations come from an elliptical distribution -- multivariate normal being the most important example of this class. The main question is if the covariances of the clusters are equal or different, and if different, what is the difference. Volume, shape and orientation are the three aspects considered. The model selection is based on the BIC, which is traditionally defined here as the expression $$B=2\log(L)-m\log(n),$$ where $L$ denotes the value of the likelihood function at the optimum, $n$ is the number of observations and $m$ the number of parameters in the model ($B$ is -1 times the "usual" BIC for regression model). So in this approach the model with the largest BIC value is chosen.

In our case the results clearly proved that the model-based approach resulted in identifiable areas as clusters, which is very much preferred, compared to the fluctuation of the simple $k$-means clustering. This has not changed, when the effect of the NAO was removed in spite of the different number of clusters needed in this case.

\subsection{Bootstrap}
\label{Ch22}

Bootstrap is a resampling method that can be used for assessing the properties of the estimators. There are many variants of the original idea of Efron. 
One approach, designed especially for assessing the reliability of regression models is the weighted bootstrap of Wu (1986). However, all these standard methods are suitable only for the case of independent 
and identically distributed errors. In case of heteroscedasticity and/or dependency among the random error terms, the standard methods fail, as it is seen by our simulation study as well as in the classical paper of Singh (1981).

The dependent weighted bootstrap is suitable for these cases by choosing a resampling  method that is capable of reproducing the dependence of the original observations. This is a relatively new concept, first introduced by Shao in 2010. The traditional tool for such cases was the block bootstrap, which itself has several variants. There are available algorithms 
for choosing the block size, which minimize the standard error of the bootstrap estimators (Politis and White, 1994). However, it is reported in Shao, 2010 that in case of missing observations, the new dependent weighted bootstrap has more favourable properties. We also check the methods in case of the linear regression,
when we assume dependency among subsequent observations. It turns out -- as shown by our simulations -- that confidence intervals, based on the proposed weighted dependent bootstrap have better coverage properties. 

In its original form the weighted bootstrap sample was constructed from the residuals of the regression model $r_i$, by multiplying them with the weights $w_i$, which were supposed to be i.i.d., with $\text{E}(w_i)=0$ and $\text{Var}(w_i)=1$. These properties ensure that for the bootstrapped residuals $w_ir_i$, we have $E(w_ir_i)=0$ and $\text{Var}(w_ir_i)=\text{Var}(r_i)$. A natural choice may be the $w_i$, for which $P(w_i=1)=P(w_i=-1)=0.5$, thus practically choosing the same residuals as observed, but with random signs. There are of course other possible choices for the distribution of $w$, for details see Wu (1986).

However, when the error terms are dependent, then the simple method seen above does not work. The bootstrap data generating process must correspond to that of the observations.
This can be achieved by the dependent weighted bootstrap, where the conditions $\text{E}(w_i)=0$ and $\text{Var}(w_i)=1$ still hold, but a dependence structure is assumed for the weights $w_i$. In 
the original paper, Shao proposed a multivariate normal distribution, with covariance matrix
$$\text{cov}(w_i,w_j)=K(\frac{i-j}{\ell}),$$
where $K$ is a kernel function and $\ell$ is a suitable norming factor (bandwidth).

In the simulation part of our paper we have first considered a small sample of $n=100$ observations, so there was no problem with defining the multivariate normal distribution via its covariance matrix. However, in the real applications we have a sequence of observations of length over 20000, so it is not feasible to generate such a long sequence of dependent normal variates this way. Thus we have also checked the properties of a simpler data generating process: namely, an AR(1) series: $w_{i+1}=rw_i+\sqrt{(1-r^2)}\varepsilon_i$, where $\varepsilon_i$ is an i.i.d. sequence of standard normal distributions. The role of $r$ is similar to that of $\ell$, they both determine the length of the memory in the sequence: a large $r$ as well as a large $\ell$ implies long range dependence. 

The theoretical properties (consistency) of the dependent weighted bootstrap were proved under somewhat strong conditions (e.g. $m$-dependency of the weights), so the cases we investigate here do not fall into the 
theoretically proven category. However, we are convinced that under suitable parametrization the method provides consistent results for much more general cases (we plan to come back to the theoretical aspects elsewhere). Our simulation results are promising, the suitably chosen AR(1) model for the weights turned out to be 
a good candidate (see Section \ref{appuni}). 

One important problem is yet to be solved: how should we choose the parameter $r$? We have faced a similar question in the paper of Rakonczai et al (2014), where the block size of the bootstrap resampling had to be found. We determined it by the best fitting AR model (VAR in the bivariate case). The fit was measured by the variance of the estimator $\bar X$ (or the trace of its covariance matrix in the bivarate case). To be more exact, in \cite{rak}, the block size was determined as the $\widehat{b}$, for which the estimated trace of the bootstrap covariance matrix was the  {nearest} to the one  derived from the fitted VAR model:
\begin{equation} \widehat{b}=\underset{1\leq b \,\in \mathbb{Z}}{\text{argmin}}
\left| 
\text{tr}\left( \text{Cov}\left( \overline{X}_{VAR} \right) \right) - 
\text{tr}\left( \text{Cov}_* (\overline{X}_b^* ) \right)
\right| ,\label{bopt}
\end{equation}
where $ \text{Cov}_* (\overline{X}_b^*)= \text{Cov} (\overline{X}_b^* | \mathcal{X}_n)$ (i.e. the covariance of the bootstrap sample based on block size $b$, under the observed sample $\mathcal{X}_n$). 
We have modified (\ref{bopt}) on a way that the parameter was the $r$ in the AR(1) model:
\begin{equation}\label{wdet}
\widehat{r}=\underset{0<r<1}{\text{argmin}}
\left| 
\text{tr}\left( \text{Var}\left( \overline{X}_{AR)} \right) \right) - 
\text{tr}\left( \text{Var}_* (\overline{X}_r^* ) \right)
\right| ,
\end{equation}
\section{Simulations}
\label{appuni}

First we checked the classical case, where the errors in the linear model are indeed independent and identically distributed. 
An independent sample from the model $y=x+\varepsilon$ was simulated, with $\varepsilon\sim N(0;1)$. The investigated methods:
\begin{enumerate}
	\item the simple Efron-type bootstrap
	\item the weighted bootstrap for the residuals
	\item the dependent weighted bootstrap with AR(1) structure for the weights and normal innovations, as $w_n=0.9w_{n-1}+\sqrt{0.19}\eta_n$, where $\eta_n$ is i.i.d. standard normal.
	\item the dependent weighted bootstrap with multivariate normal weights, the covariance was given by $\sigma_{i,j}=1-|i-j|/\ell$ for $\ell=25$.
\end{enumerate}
Here the results show that all of the investigated bootstrap methods are near to unbiasedness -- considering the coefficients of the linear regression. However the dependent weighted methods have understandably larger variance, which underlines that their application is unnecessary, if independence can be accepted -- as e.g. the confidence intervals will not have the desired coverage in these cases.

Next we have simulated sequences of length $n=23360$ (the number of observations in our data set) with AR(1) structure for the residuals with $r=0.812$ (the average of our estimators) and a trend coefficient of $8.6*10^{-5}$ (again a typical value for our data set). The repetition size was 500, and we compared the performance of the three bootstrap methods, proposed to cases like ours. The results are shown in the Table \ref{boott}, containing $10^5$-times the estimated values. Here we can observe the superiority of the weighted bootstrap. The block size for the block bootstrap
was determined by the Politis-White algorithm.

\begin{table}[!h]
	\begin{center}
		\begin{tabular}{|l||r|r|r|r|r|r|r|} 
						\hline
			method \hspace{1cm}   quantile & 0.025 & 0.05& 0.25& 0.5& 0.75& 0.95 & 0.975
			\\ \hline  \hline 
			\bf {AR(1) process} &6.64& 6.95&8.05& 8.53& 9.12& 10.11& 10.43\\ \hline  
			block bootstrap  &5.34& 5.64& 7.72& 8.86& 10.39& 12.19& 		12.62\\ \hline 
			indep. weighted boot& 7.62& 7.75& 8.30& 8.61& 8.94& 9.41&		9.50 \\ \hline 
			dep. weighted boot&6.74& 7.14& 8.13& 8.77& 9.40& 10.22&	10.47	\\ \hline 
		\end{tabular}
	\end{center}
	\caption{The effect of bootstrap type on the estimated confidence interval for the trend coefficient in case of normally distributed AR(1) error structure with $r=0.812$. The block bootstrap is 
		too conservative with extreme quantiles; the independent weighted bootstrap on the other hand is too optimistic. The dependent weighted bootstrap (with the AR(1) dependence, determined by (\ref{wdet}) turned out to be the best. The number of repetitions was $n=500$.}
	\label{boott}
\end{table}

\begin{figure}[ht]
	\centering
	\includegraphics[height=70mm]{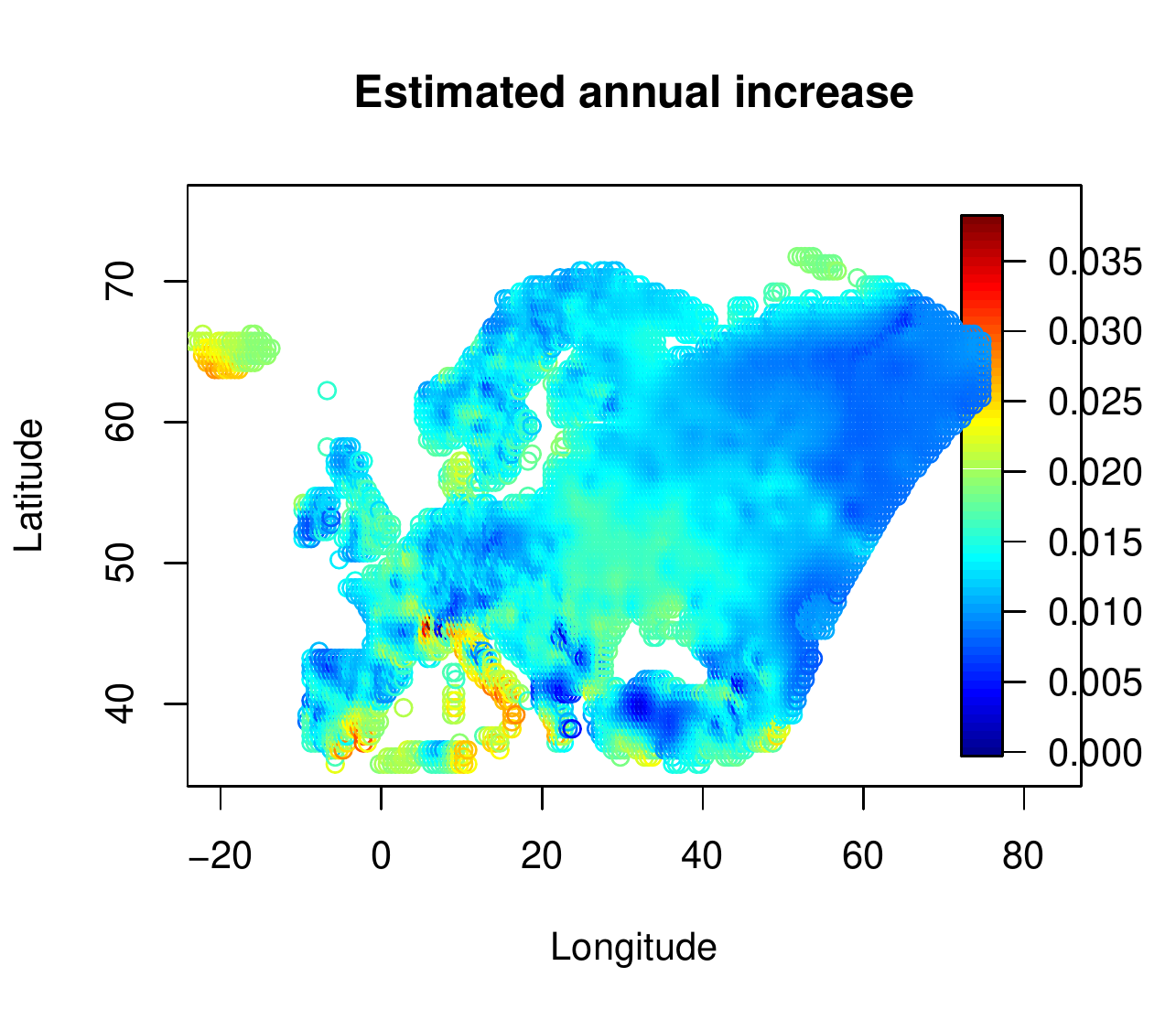} 
	
	\caption{Maximum of the regression coefficients for the grid points}
	\label{max}
\end{figure}

\section{Applications}
\label{appl}
Having shown the tools, we utilize them in the data analysis.

We have analysed 65 years of daily temperature data of the 0.5 grade European grid points, shown on Figure \ref{map}.
We have not investigated the time series in detail, but it is obvious that seasonality is its most important feature. So first we have standardized the data for every day of the year 
 by simple nonparametric polynomial (Loess) smoothing, using both first- and second-order standardisation -- mean and variance were thus approximately constant within the year.  

This standardized daily data was used as a basis for the simple linear regression: $x_t=at+b$. We are interested in the steepness of the regression line (measured by the estimated coefficient $\hat a$), as well as in the strength of the model (as shown by the coefficient $R^2$). These parameters were calculated for the first 30 years (in the $k$th equation we consider data from year $k$ till 65).

The maximum value of the estimated regression coefficients is shown for each grid points in Figure \ref{max}.

\begin{figure}[!ht]
	\centering
	\includegraphics[height=70mm]{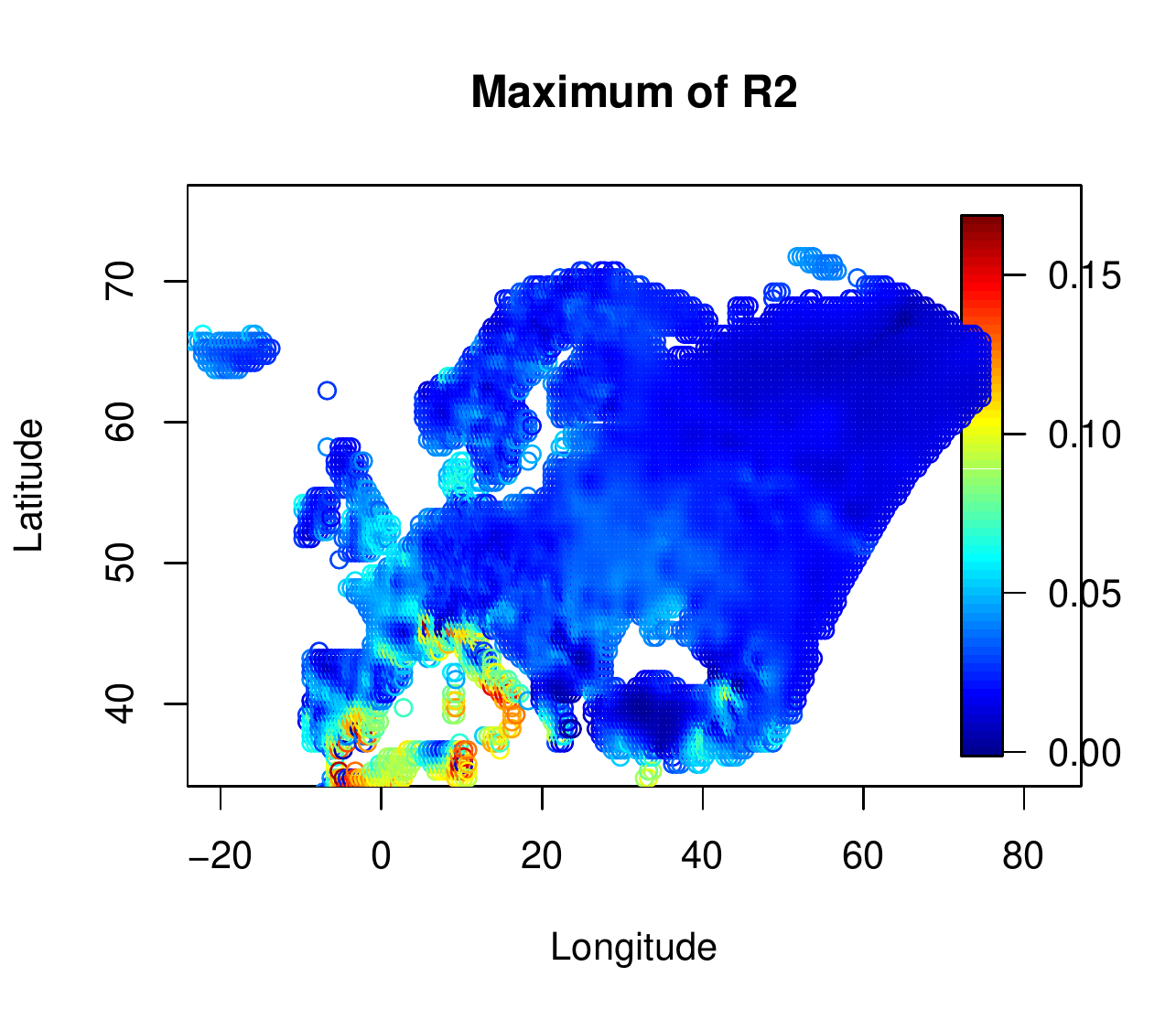} 
	
	\caption{$R^2$ values of the regression coefficients for the grid points}
	\label{r2}
\end{figure}

A similar plot for the strength of the model is Figure \ref{r2}. Not surprisingly, we see the larger values mostly in the regions with higher coefficients. The low values of $R^2$ are quite natural, as there are many more nonlinear disturbances in the weather, compared to the slow but steady global warming.

\begin{figure}[!ht]
	\centering
	\includegraphics[height=70mm]{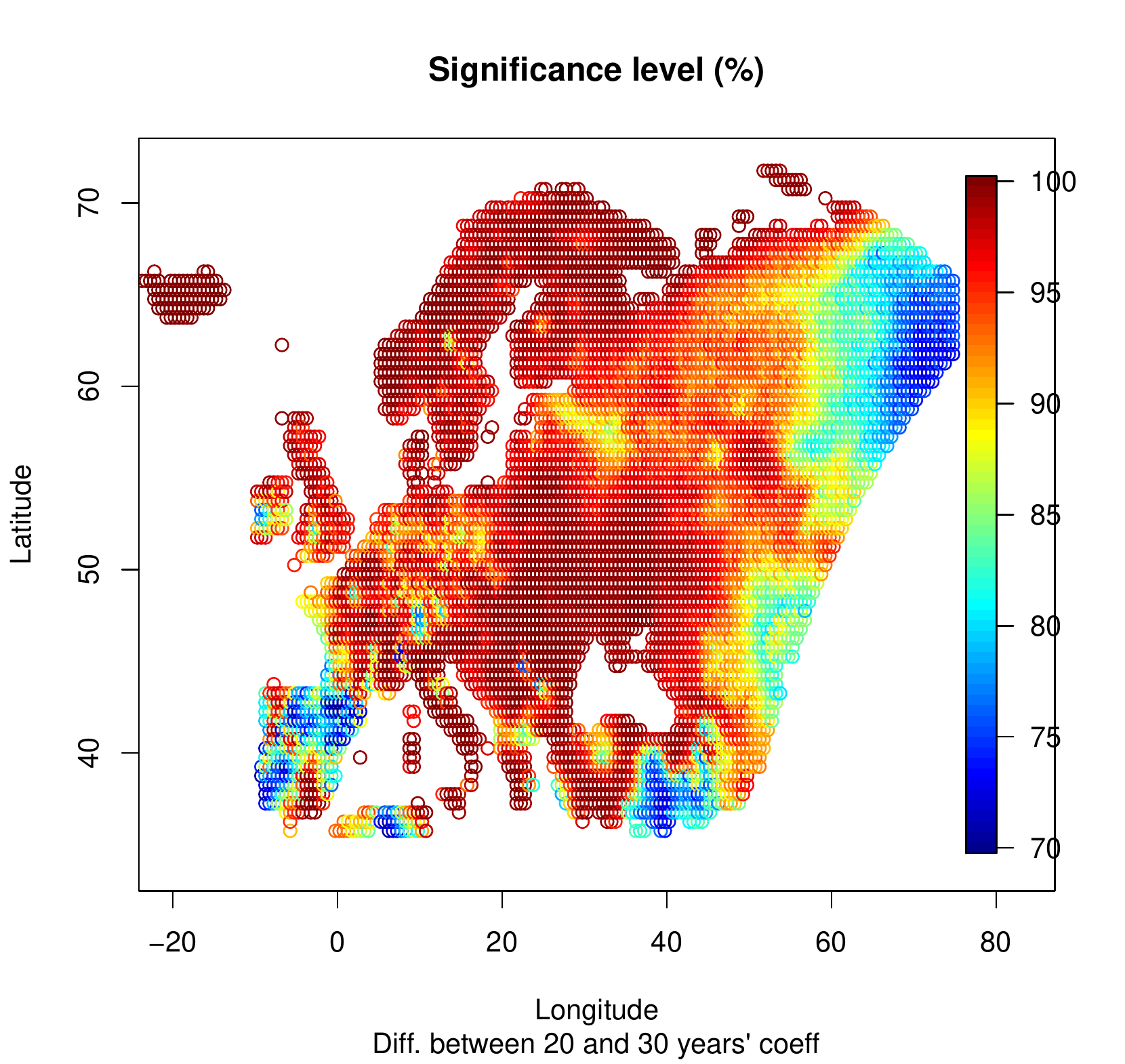} 
	
	\caption{Significance of the increase of warming, when the last 35 years ($k=30$) is compared to the last 45 years ($k=20$)}
	\label{sig}
\end{figure}

One interesting question is the significance of the results for the climate data, we have analysed -- especially as we have considered data segments of different lengths.

So we performed a small study, were the linear coefficient was defined as $10^{-4}$ and a simple AR(1) structure for the residuals was assumed: $\varepsilon_n=0.9\varepsilon_{n-1}+\sqrt{0.19}\eta_n$, where the sequence $\eta_n$ was i.i.d. standard normal. Here we have used the dependent weighted  bootstrap method introduced in Subsection \ref{Ch22}, and we focused on the steepness coefficient of the linear model. It turned out that the estimators like this are not significant for much shorter series, but the case of $n\ge 30*365$ turned out to be sufficient for $\alpha=0.05\%$ (see Table \ref{b0}). We have pursued this idea further by applying the dependent weighted bootstrap method for our (NAO-adjusted) data. Having repeated the simulations 100 times for each data point for $k=20$ and $k=30$, we have got 2 times 100 coefficients for the grid points (let us denote them by $x_{20}$ and $x_{30}$). We may estimate the significance of the increase of the coefficient by calculating the percentage of the pairs where  $x_{30}$ is larger than  $x_{20}$. These values are plotted on Figure \ref{sig}. One may see that the increase is highly significant for Scandinavia and large parts of central Europe.

\begin{figure}[!ht]
	\centering
	\includegraphics[height=60mm]{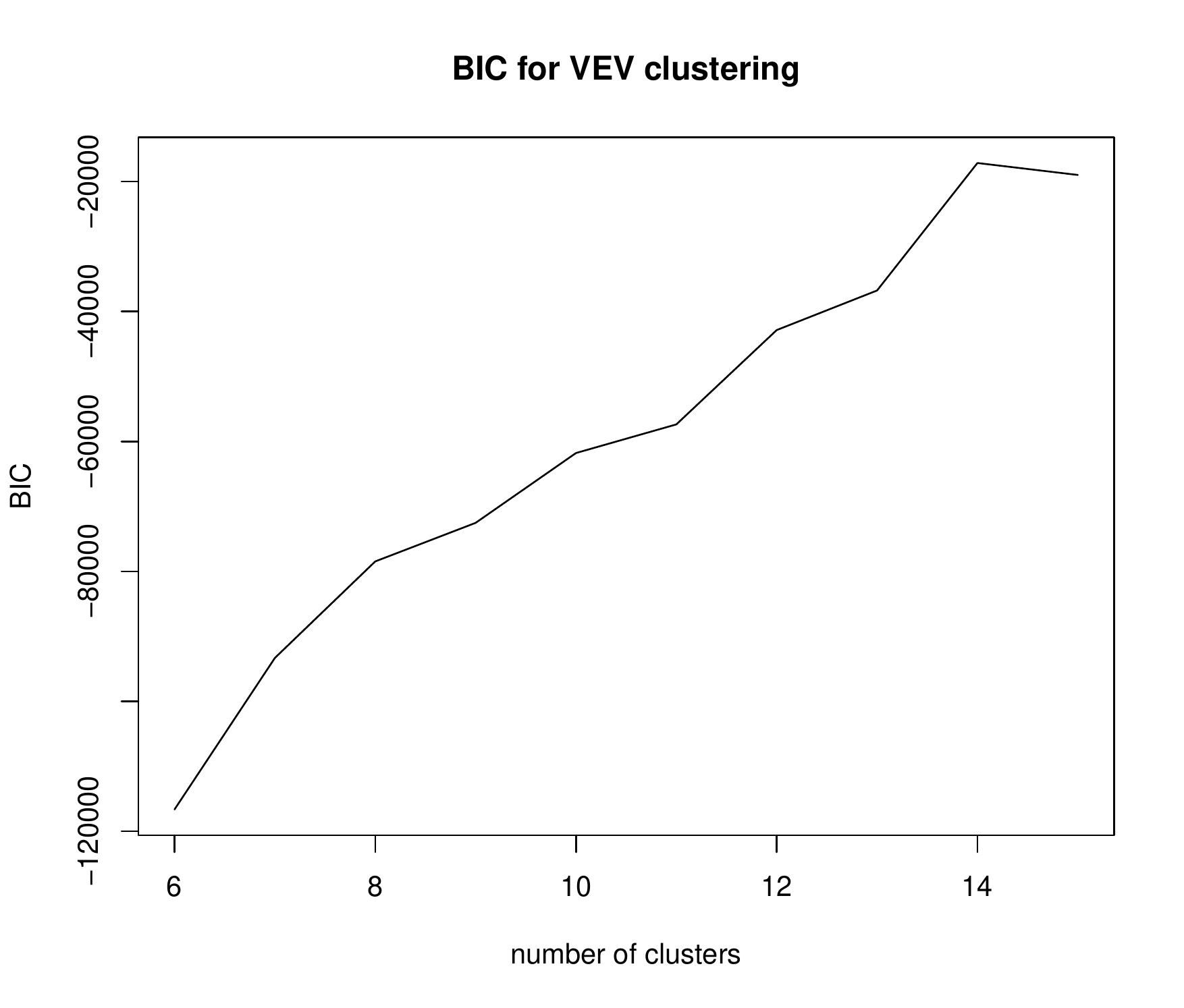} 
	
	\caption{BIC values for the model-based clustering}
	\label{BIC}
\end{figure}

\begin{table}[!h]
	\begin{center}
		\begin{tabular}{|l||c|c|c|} 
			   \hline
			\,\, number of years & 10 & 30& 60
			\\ \hline  \hline 
			proportion of negative values (\%) &26.6& 0.17& 0	\\ \hline 
		\end{tabular}
	\end{center}
	\caption{A simulation for a typical sample point. We assumed an AR(1) error structure with the estimated coefficients. The weighted bootstrap was applied 500 times for a given sequence, 100 such sequences were investigated. The values show the percentage of negative estimated coefficients. }
	\label{b0}
\end{table}

We have performed a model-based $k$-means clustering of the grid points for the regression coefficients. Similar approach was used for a completely different data in  \cite{hz}. The clearly preferred model was the so-called VEV (ellipsoidal covariance structure with equal shape). The analysis of the BIC-values (Figure \ref{BIC}) shows that 14 clusters should be formed.

\begin{figure}[!ht]
		\centering
	\includegraphics[height=70mm]{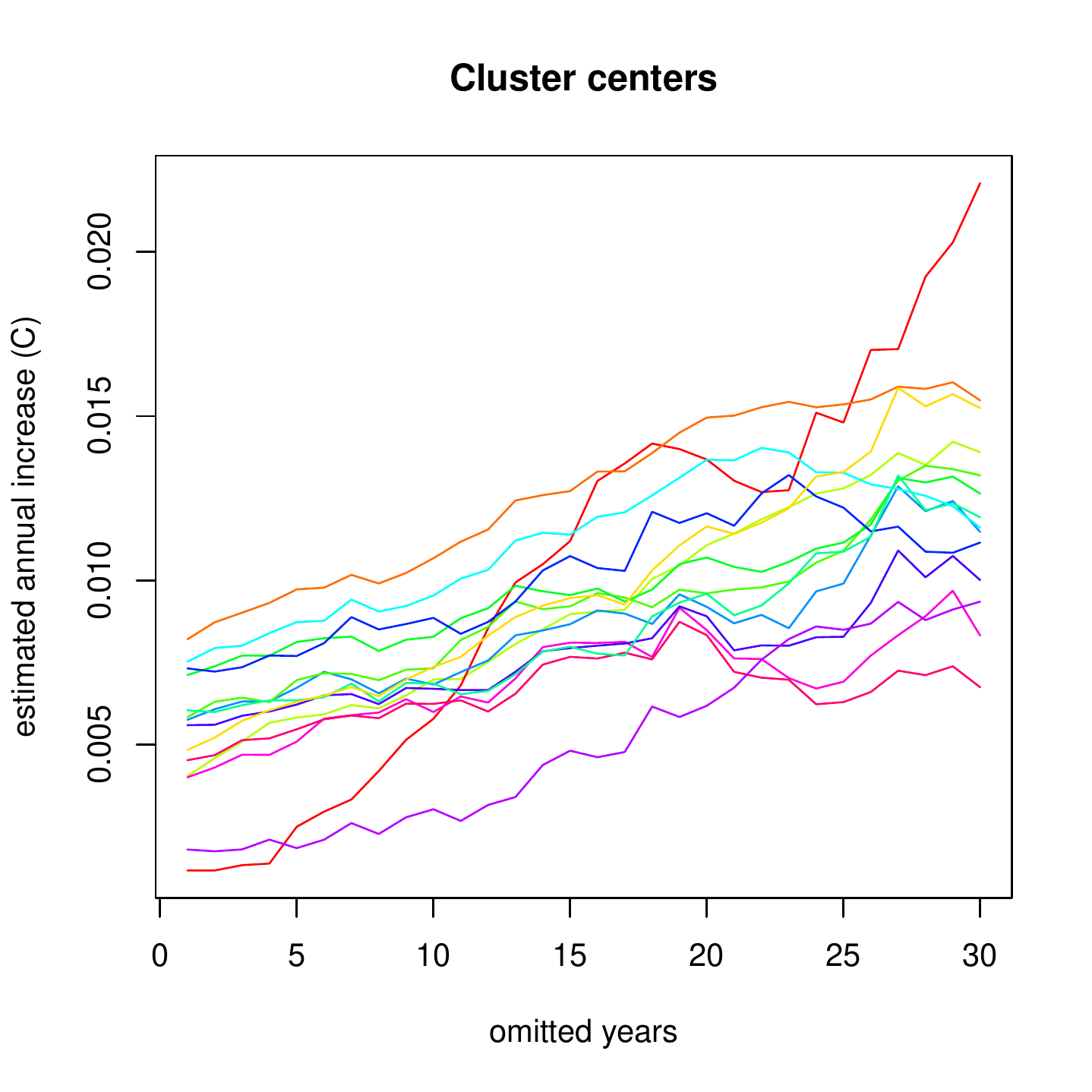} 
	
	\caption{Cluster centers, based on the regression coefficients for the grid points (14 clusters, based on the Gaussian mixture method)}                                                                                                                                                                          	\label{clusc}                                                                                                                                                    \end{figure}                                                                                          
\begin{figure}[!ht]
		\centering
	\includegraphics[height=70mm]{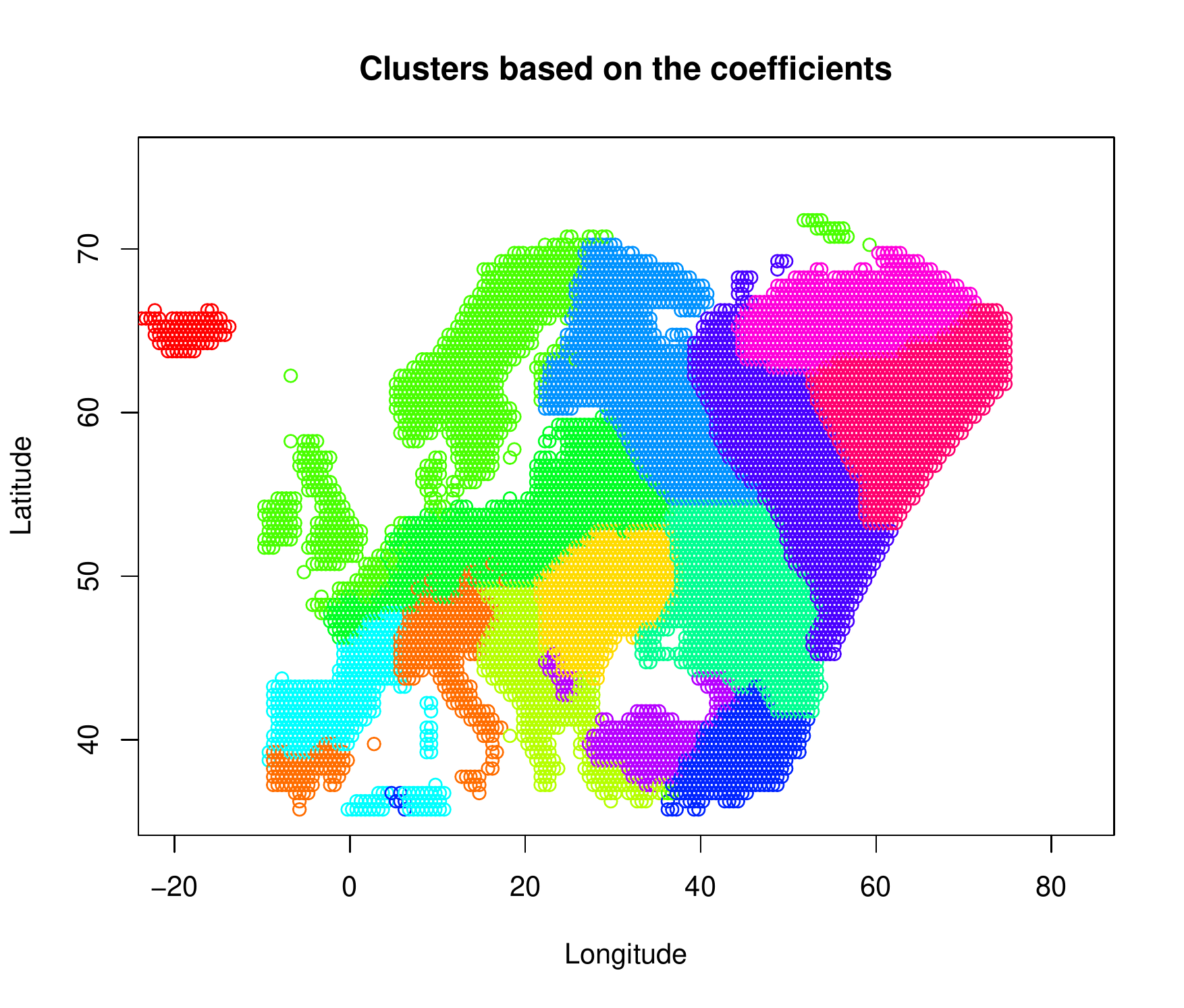}                                                                                                                                                                 	
	\caption{Clustering, based on the maximum of the regression coefficients for the grid points (14 clusters, based on the Gaussian mixture method)}
	\label{c4}
\end{figure}                                                                                                                                                                                                                                                 
Figure \ref{clusc} shows the time-development of the 14 cluster centers. Almost all of the curves show a clear upward trend, i.e. in almost all cases the coefficient increases as the number of omitted years from the beginning of the investigated period increase. This means high increase in the last years, in accordance with the widely accepted phenomenon of the global warming. However the coefficients (i.e. the speed of the temperature increase) are different in the clusters. 
The  clusters were differing mainly in baseline level, the lines were otherwise quite parallel.
The clusters are well separated. It is of definite meteorological interest to investigate which areas belong to the clusters. The colors of Figure \ref{c4} correspond to those of Figure \ref{clusc}, so we may state that large parts of Central Europe belong to the yellow cluster (which is among the regions with the highest annual increase). The highest estimated increase is observed in the Western part of the Mediterranean region and on Iceland while the European part of Russia showed the slowest increase. The quick, further temperature-increase in the already warm regions might be interesting from a medical point of view as well, since the further warming of this region might result in quicker than expected outbreaks of tropical epidemics like malaria.

\begin{figure}[!ht]
	\centering
	\includegraphics[height=70mm]{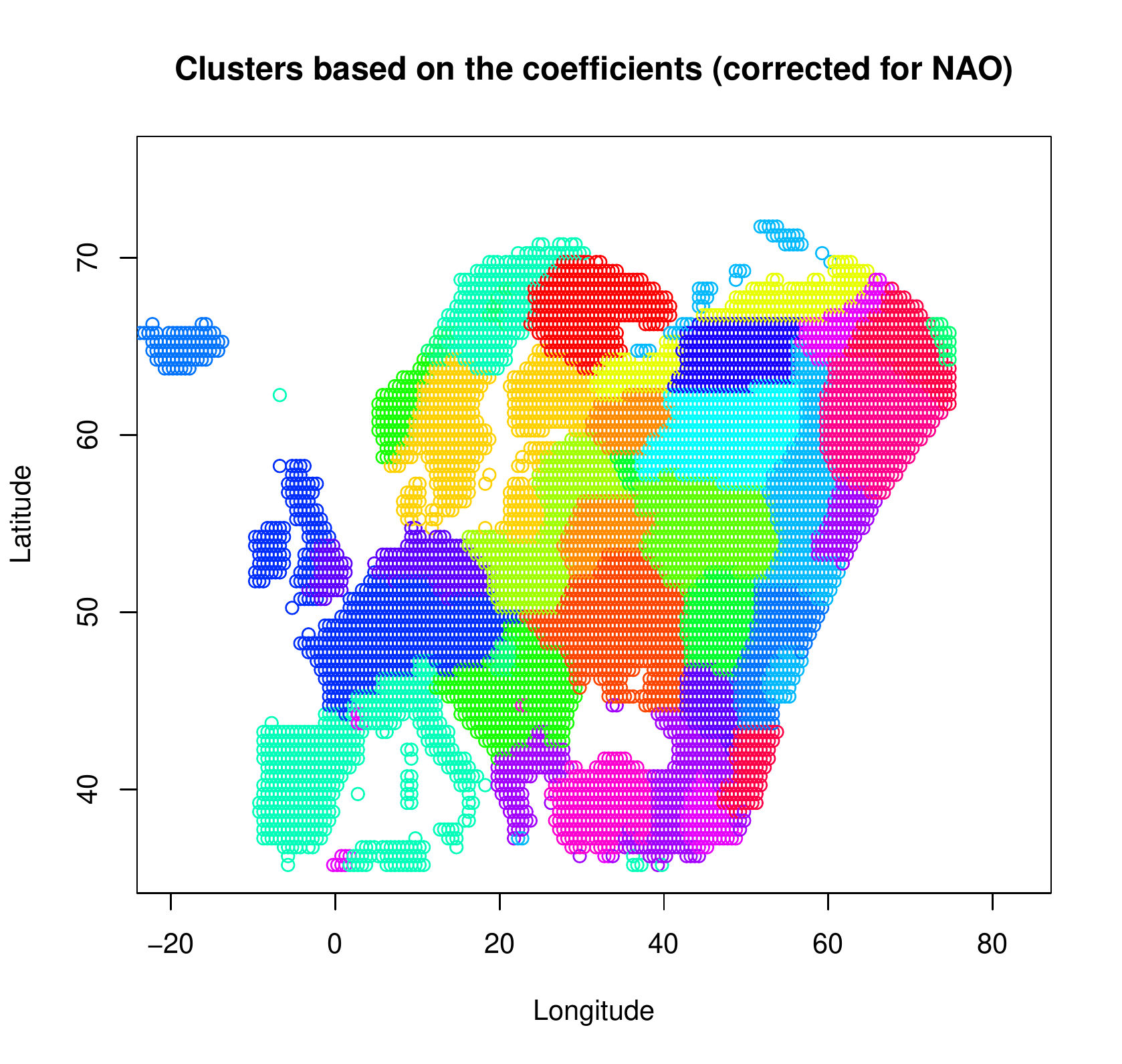}                                                                                                                                                                 	
	\caption{Clustering of the maximum of the regression coefficients for the NAO-corrected data (22 clusters, based on the Gaussian mixture method)}
	\label{na_clu1}
\end{figure}   

\begin{figure}[!ht]
	\centering
	\includegraphics[height=70mm]{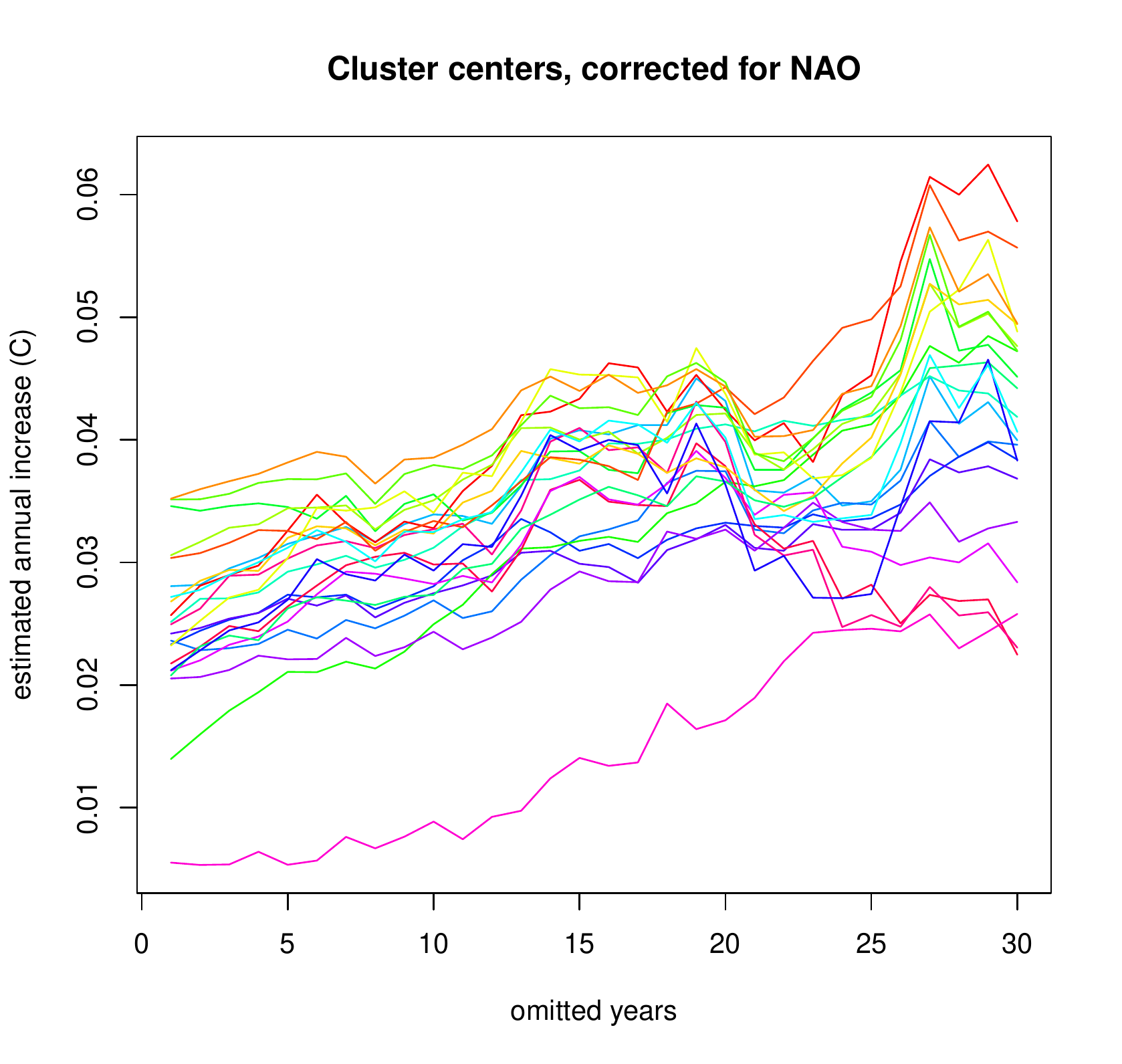}                                                                                                                                                                 	
	\caption{The cluster centers of the maximum of the regression coefficients for the NAO-corrected data (22 clusters, based on the Gaussian mixture method)}
	\label{na_clu2}
\end{figure}   

We have repeated the above procedure for the NAO-adjusted data. It gave similar results, with well separated clusters. The results are shown in Figures \ref{na_clu1} and \ref{na_clu2}.
Here more (22) clusters were suggested by the procedure above, but the spatial pattern of the clusters is still plainly visible, and the behaviour of most of the centers has not changed either. When running the dependent weighted bootstrap simulation, we have got that e.g. for $k=20$ only 1.5\% of the individual estimators are non significant. These mostly belong to the cluster with the exceptionally low values on Figure \ref{na_clu2}, in Turkey. 

Our results, shown on Figure \ref{sig} are in accordance with  Figure \ref{na_clu2}, proving that in most of the cases this increase between points 20 and 30 is indeed significant at the 95\% level.


\section {Conclusions} 
\label{conclu}

As a conclusion we can claim that to analyse the temperature data by focusing on the last years was a sound idea, as we found interesting patterns in the gridded temperature data. The Gaussian model based clustering has resulted in a clear pattern of different regions, which might be a useful start for further climatic research. The inclusion of the NAO index showed that the results are robust enough under additional assumptions on e.g. low frequency variability.

The bootstrap is indeed an important tool in evaluating the significance of our results. However, one has to be aware of its properties. In case of dependent data, we have to take this dependence into account, when planning the bootstrap data generating process. We have compared the available methods and it turned out that the dependent weighted bootstrap is the most accurate in our case, when the dependency is simply modelled by an AR(1) process. We have also presented a practical method for determining the strength of the dependence -- which is analogous to the approach, based on the effective sample size. It has proved that the found positive coefficients are indeed significant for most of the chosen time intervals and almost all grid points. Furthermore, it has turned out that the increase in the speed of the global warming is significant even in a range of 10 years as well, for large parts of Europe, ensuring that the threat of uncontrolled warming is indeed a real threat for most of Europe. The dependent weighted bootstrap thus has proven to be a unique tool for checking hypotheses about regression models.


\section*{Acknowledgment}

We acknowledge the E-OBS dataset from the EU-FP6 project ENSEMBLES (\url{http://ensembles-eu.metoffice.com}) and the data providers in the ECA\&D project (\url{http://www.ecad.eu}).

The research of A. Zempl\'eni was supported by the Hungarian National Science Foundation (OTKA, K-81403).



\end{document}